# Scalable Extraction Based Semantic Communication for 6G Wireless Networks

Yuzhou Fu, *Student Member*, *IEEE*, Wenchi Cheng, *Senior Member*, *IEEE*, Wei Zhang, *Fellow*, *IEEE*, and Jingqing Wang, *Member*, *IEEE*

*Abstract*—Due to the challenges of satisfying the demands for communication efficiency and intelligent connectivity, sixth-generation (6G) wireless network requires new communication frameworks to enable effective information exchange and the integrated Artificial Intelligence (AI) and communication. The Deep Learning (DL) based semantic communication, which can integrate application requirements and the data meanings into data processing and transmission, is expected to become a new paradigm in 6G wireless networks. However, existing semantic communications frameworks rely on sending full semantic feature, which can maximize the semantic fidelity but fail to achieve the efficient semantic communications. In this article, we introduce a novel Scalable Extraction based Semantic Communication (SE-SC) model to support the potential applications in 6G wireless networks and then analyze its feasibility. Then, we propose a promising the SE-SC framework to highlight the potentials of SE-SC model in 6G wireless networks. Numerical results show that our proposed SE-SC scheme can offer an identical Quality of Service (QoS) for the downstream task with much fewer transmission symbols than the full semantic feature transmission and the traditional codec scheme. Finally, we discuss several challenges for further investigating the scalable extraction based semantic communications.

## I. Introduction

As we enter the era of intelligent interconnection, the widely deployed devices need to support various emerging wireless applications, which poses two main challenges in developing sixth-generation (6G) wireless networks. First, the potential applications in 6G wireless networks demand improved communication efficiency, where these applications encompass a wide range of domains, including extended reality, intelligent transportation, holographic telepresence, collaborative robots, and so on. Second, the International Telecommunication Union (ITU) 2030 group published the latest vision on 6G architecture, identifying the integrated AI and communication as a new scenario to provide services beyond communications [1]. Therefore, 6G wireless networks will incorporate AI from the very beginning of its design to integrate the meaning of information into data processing and transmission, thus substantially increasing network robustness, network performance, and communication efficiency. However, the current wireless communication framework fails to utilize the meaning of messages during data processing and transmission, it becomes the major bottleneck for developing 6G wireless networks. Essentially, the bottleneck results from the hypothesis of Shannon's information theory that is applied in the physical layer and considers information to be statistically independent at the bit level. This hypothesis neglects the meaning of information and the knowledge accumulated during the communication, thereby impeding the integration of data transmission and the meaning exchange.

Semantic communication, initially introduced by Weaver [2], as an extension of Shannon theory, focuses on how to guide the receiver towards intended goal by conveying the meaning of data. The deep learning (DL) based semantic communication, which serves as an instantiation of classic semantic communication, can compress the transmission data to the compact feature representations (namely, semantic features) carrying local meanings as well as context information. By conveying the semantic features of data, the DL based semantic communication can offer better Quality of Service (QoS) for the downstream task than the traditional communication but requires much fewer transmission symbols. Thus, the DL based semantic communication is expected to enable the integration of AI and communication in 6G wireless networks and support the the potential applications in 6G wireless networks [3].

Recently, many DL based semantic communications frameworks have been proposed for wireless communications [4]–[6]. The authors of [4] introduced the attention mechanism into end-to-end semantic communication to extract the features of speech signal, where the semantic communication aims to minimize the mean-square errors of speech signal. In [5], the authors proposed DL based Joint Source-Channel Coding (JSCC) for conveying the features of text over noise channels. For the image transmission, the authors of [6] mapped the image data to the image features and learns noise resilient coded representations for conveying the semantic feature, thus outperforming the traditional digital communication at noise channels.

Although the feasibility of DL based semantic communication is validated, the research on the efficient semantic based wireless communications is still in its infancy. The existing semantic communications frameworks rely on sending

This work was supported in part by the National Key R&D Program of China under Grant 2021YFC3002102, in part by the Key R&D Plan of Shaanxi Province under Grant 2022ZDLGY05-09, and in part by the Key Area R&D Program of Guangdong Province under Grant 2020B0101110003. (*Corresponding author: Wenchi Cheng.*)

Yuzhou Fu, Wenchi Cheng, and Jingqing Wang are with State Key Laboratory of Integrated Services Networks, Xidian University, Xi'an, 710071, China (e-mails: fyzhouxd@stu.xidian.edu.cn; wccheng@xidian.edu.cn; jqwangxd@xidian.edu.cn).

W. Zhang is with School of Electrical Engineering and Telecommunications, The University of New South Wales, Sydney, NSW 2052, Australia (e-mail: w.zhang@unsw.edu.au).

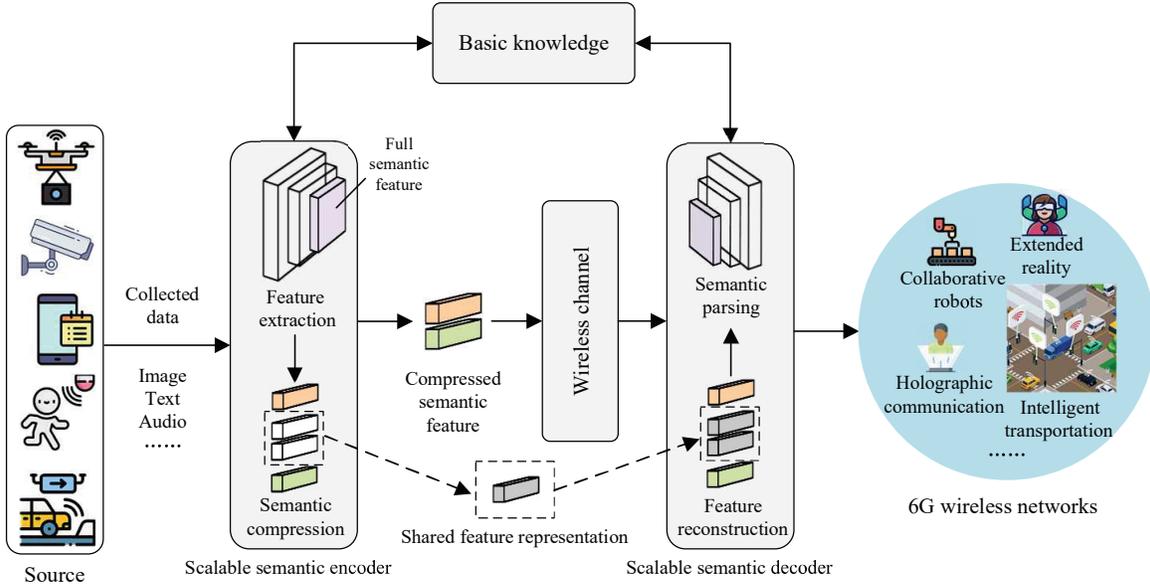

Fig. 1: The scalable extraction based semantic communication model for 6G wireless networks.

the full semantic feature, which can maximize the semantic fidelity of the data but fail to achieve the efficient semantic communications. In practice, the semantic communications can restore the meaning of original data according to the prior information without receiving full semantic feature [2]. Also, for the downstream tasks (e.g., object detection, data reconstruction, and semantic segmentation) driven by the semantic feature, the full semantic feature extracted from original data contain a large amount of redundant features [7], which is useless for the execution of downstream tasks. As a result, the efficient semantic communication can increase the communication efficiency by further compressing redundant semantic feature. Motivated by the development of efficient semantic communications for 6G wireless networks, we plan to exploit a novel scalable extraction based semantic communication framework, which can offer an identical QoS for the downstream tasks as the full semantic feature transmission by focusing on the Features of Interest (FoI) of the receiver. In particular, the FoI refers to the semantic features most relevant to the downstream task and will not be compressed during transmission.

## II. THE SCALABLE EXTRACTION BASED SEMANTIC COMMUNICATION MODEL

As shown in Fig. 1, we establish the scalable semantic communication model to support the potential applications in 6G wireless networks. To guarantee the ideal QoS for users, these applications need to strictly satisfy the end-to-end latency and data rates requirements. To this end, existing semantic communications can map the data to the semantic features, thus compressing the redundant information in the data. In addition to the compression of the data, the SE-SC model aims to further compress redundant semantic features and only send the necessary semantic features for the application to offer an identical QoS for the application using much fewer transmission symbols. Our proposed SE-SC model consists of the scalable semantic encoder, the scalable semantic decoder, and the basic knowledge. The basic knowledge is the representation of knowledge associated with the semantics of the data. This knowledge representation, which is the basis for AI mechanisms with semantic analysis ability, is used to efficiently represent and interpret data by defining the semantic feature carried by the data. However, the basic knowledge has not been fully used in existing semantic communications.

More specifically, the scalable semantic encoder aims to send the compressed semantic feature, which further filters out redundant information but will not reduce the QoS of the applications. The encoding process is made of the feature extraction and the feature compression. In particular, the feature compression is responsible for compressing the redundant semantic features in the full semantic feature. The redundant semantic features refers to the semantic feature that are easily predicted based on context or are useless for driving the downstream task. During feature compression, we train a feature representation shared by the transmitter and the receiver to represent redundant features, and thus it is a lossy feature compression. From the perspective of classic lossy compression, this data processing will lead to data distortion. However, due to the introduction of basic knowledge, the performance of scalable communication remains unaffected by this process since the required data meanings for the application can be perfectly recovered. On the other hand, the scalable semantic decoder, which consists of two basic parts including the feature reconstruction and the semantic parsing, is responsible for reconstructing required data for the downstream tasks based on the received semantic feature.

## III. THE FEASIBILITY OF THE PROPOSED SE-SC MODEL

In order to verity the the feasibility of the SE-SC model, we conduct a simulation based on compressed semantic feature serial number of unsent semantic features. In practice, we can compress any number of semantic features into a shared feature representation thus achieving different semantic feature compression ratio. The semantic similarity is a positive metric for the semantic communications and can be calculated based on the semantic feature space [5], [12], [13]. The semantic feature space is mapped from the input data using the neural network based tool. Such space amplifies the distance of semantically inconsistent data while reducing the distance of semantically consistent data.

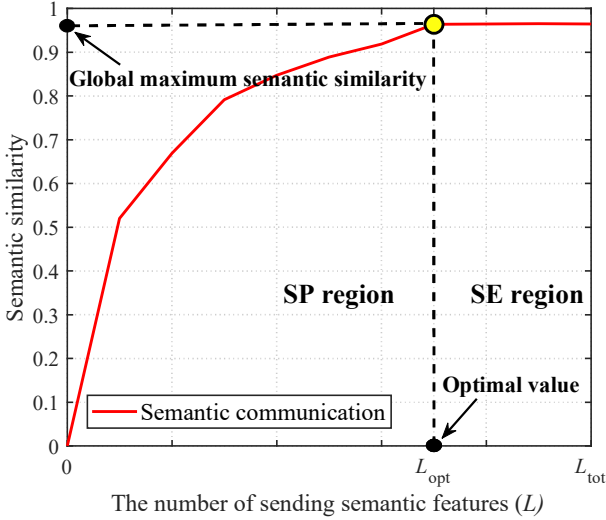

Fig. 2: The relationship between semantic similarity and the number of sending semantic features.

To analyze the semantic similarity, the simulation results as shown in Fig. 2 reveal the relationship between the semantic similarity and the number of sending semantic features, where $L$ is the number of sending semantic features, $L_{\mathrm{opt}}$ denotes the optimal value and $L_{\mathrm{tot}}$ represents the total number of semantic feature corresponding to the transmission data. In Fig. 2, we define Semantic Parsing (SP) region and Semantic Explicit (SE) region, which are divided by the vertical dashline in Fig. 2, to determine whether the goal of semantic communication is to achieve the intended semantic similarity or maximize the semantic similarity. It can be observed from Fig. 2 that the semantic similarity increases as $L$ increases in the SP region ($0 \leq L \leq L_{\mathrm{opt}}$). While in the SE region ($L_{\mathrm{opt}} < L \leq L_{\mathrm{tot}}$), the semantic similarity converges to a fixed value when $L$ is larger than $L_{\mathrm{opt}}$. This is because when $L$ reaches $L_{\mathrm{opt}}$, the semantic communication can successfully restore the original meaning of information according to the basic knowledge, thus maximizing the semantic similarity. In the SP region, we can control the number of sending semantic features to achieve the intended semantic similarity for supporting the corresponding downstream task. In the SE region, we can find the optimal number of sending semantic features and only transmission under slow Rayleigh fading channel, where the receiver aims to reconstruct data according to the received semantic feature. The simulation results reveal the relationship between the semantic similarity and the number of sending semantic features. In this simulation, the SE-SC model is based on the Masked AutoEncoders (MAE) [10]. In order to obtain the robustness of channel noise, we introduce the power normalization layer and non-trainable wireless channel model during training phase [5]. Also, a selected subset of the semantic features and a binary sequence are sent to the receiver, where the binary sequence is used to flag the send the corresponding number of semantic features to achieve maximum semantic similarity while saving communication resource. Thus, it is desirable to exploit efficient semantic based wireless communication framework, by adaptively optimizing the number of transmission semantic features.

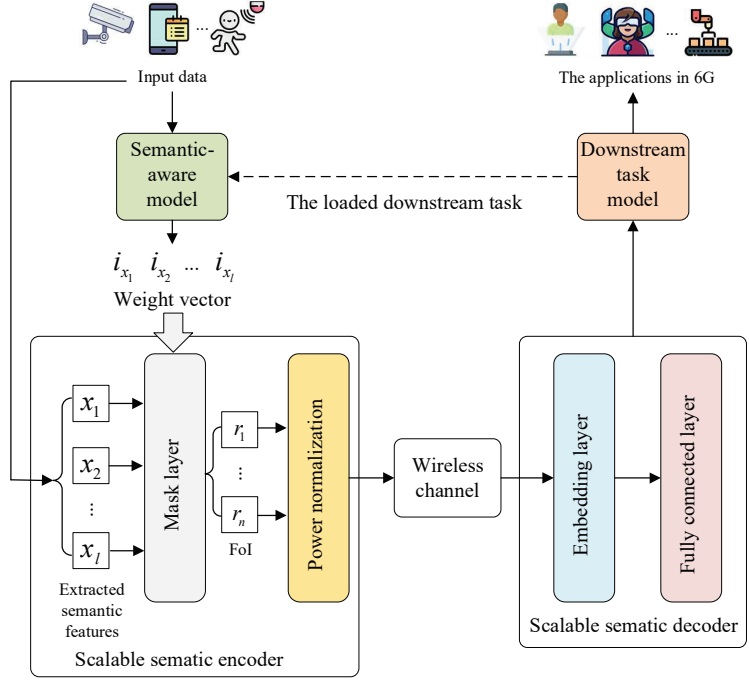

Fig. 3: The scalable extraction based semantic communication framework in 6G.

TABLE I: Some Alternative Neural Network Architectures for Semantic-Aware Model and Scalable Semantic Codec.

| Backbone module | NN architecture | Benefits | Limitations |
| --- | --- | --- | --- |
| Semantic-aware model | With the intermediate feature mappings of the downstream task model as input [8], the semantic-aware model can produce the weight vector using the Convolutional Neural Networks (CNNs). | This scheme is suited for producing the weight vector of various types of data, where the weight vector can directly reflect the semantic importance of each semantic feature for the downstream task model. | The semantic-aware model need to load corresponding downstream task model for extracting the intermediate feature mappings as input parameter, resulting in high computing complexity of semantic-aware model. |
| | An entropy model can be used to evaluate the entropy of each semantic feature in order to produce the weight vector, where semantic feature with high entropy means important semantic information [9]. | The entropy model is with low complexity and is easy to be integrated into the semantic encoder, thus enabling efficient model training. | The entropy model is successfully applied to the data reconstruction, but the feasibility and effectiveness of this scheme are yet to be verified for other downstream tasks. |
| Scalable semantic codec | Based on the masked transformer model [10], the scalable semantic codec can not only learn a shared vector as shared feature representation, but also used for reconstructing intended data according to a subset of semantic feature thus achieving scalable semantic transmission. | The masked transformer model is based on the self-attention mechanism resulting in excellent semantic extraction capacity. Also, the transformer model can be used to process various types of data, such as image [10] and text [5]. | The computing complexity of the masked transformer model is high, which causes difficulties in training the semantic codec. |
| | The masked CNN based scalable semantic codec can integrated the quantizer and the entropy coding into the semantic encoder, thus enabling feature compression as well as adaptive rate transmission [11]. | Compared with the masked transformer model, the masked CNN is with lower network complexity and can achieve more flexible scalable semantic extraction based on the quantizer. | The masked CNNs is the lack of semantic processing capability for text data. |

## IV. THE FRAMEWORK OF SCALABLE EXTRACTION BASED SEMANTIC COMMUNICATION

Figure 3 illustrates our proposed SE-SC framework, which aims to send the Features of Interest (FoI) to the downstream task model for supporting the applications in 6G. In particular, the FoI refers to the semantic features most relevant to the downstream task and will not be compressed during transmission. As shown in Fig. 3, the semantic features extracted from the input data are defined as $(x_1, x_2, ...x_l)$, where $l$ represents the total number of semantic feature. Motivated by that different regions of the input data vary in semantic importance for driving the downstream task, an AI based semantic-aware model is designed to produce the weight vector based on the loaded downstream task as well as the input data, where the weight vector is used to determine the importance of each semantic feature in the downstream task model. The weight vector is defined as $(i_{x_1}, i_{x_2}, ...i_{x_l})$, where $i_{x_l} \in [0, 1]$ is the weighted value of $x_l$ and is with higher values representing more importance for the downstream task. Under the guidance of the weight vector, the mask layer can abandon the redundancy semantic features by using mask operation while preserving the FoI, thus achieving intelligent feature compression. Let $(r_1, r_2, ...x_n)$ be the FoI, where $n$ represents the total number of the FoI, then $n < l$. In particular, the masking is widely used in pre-training phase but is not yet used to develop efficient semantic communications. For wireless transmission, the power normalization layer is used to map the compressed semantic feature to the channel input sequences. During training phase, a series of non-trainable layers are used to model the widely used wireless channel models, thus enabling end-to-end communication framework. At the scalable semantic decoder, an embedding layer can reconstruct the full semantic feature based on the received semantic feature as well as learned vector. Then, a fully connected layer is used to map the reconstructed semantic feature to the space of original data and then export the output data to drive downstream task. In practice, our proposed SE-SC framework can be deployed to the air interface proposed by the 3rd Generation Partnership Project (3GPP). Specifically, the semantic-aware model is deployed in the data link layer to produce the weight vector before splitting and encoding transmission data. Also, the scalable semantic codec is deployed in the physical layer to enable adaptive coding that matches the transmission rate under guidance of the weight vector.

As the backbone module of the SE-SC framework, some alternative Neural Network (NN) architectures for the semantic-aware model and the scalable semantic codec are shown in Table I. Different from the existing semantic communications frameworks, the SE-SC framework relies on the semantic-aware model and the scalable semantic codec to achieve efficient as well as intelligent semantic communication. For the semantic-aware model, the intermediate feature mapping can more intelligently enable semantic-aware compression and transmission as compared with the entropy model, but increase the computing complexity of the semantic-aware model. For the scalable semantic codec, the masked transformer model is expected to support scalable semantic extraction of various types of data. Also, the masked CNN is widely used in the semantic extraction of image data and speech signals. The masked CNNs based scalable semantic codec is with low network complexity and thus can be applied to the wireless mobile devices. Based on the SE-SC, some basic advantages are discussed in detail below.

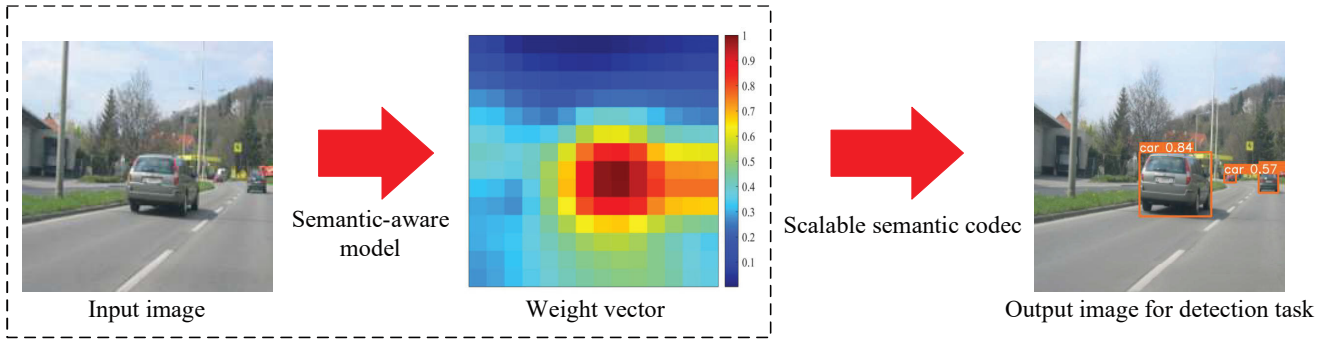

Fig. 4: A visual example of our proposed SE-SC framework in object detection task.

**Advantage 1: High Code Efficiency**: The SE-SC aims to only send compressed semantic feature instead of conveying full semantic feature and can fully use the basic knowledge of information to achieve the intended communication goal. Thus, the SE-SC can achieve higher code efficiency than the existing semantic communication. Considering the staggering amount of data in future 6G wireless networks, the SE-SC is expected to support high data rate transmission without consuming more code-domain resources.

**Advantage 2: Wide Range of Versatility**: In future 6G wireless networks, the wireless communication scenarios can be classified into three categories as human to human communication, human to machine communication, and machine to machine communication. The SE-SC is able to send the most suitable semantic feature based on the semantic-aware model. For human-to-human communication scenario, the SE-SC relies on the human cognitive ability to compress semantic feature. For human-to-machine communication scenario, the goal of SE-SC is to increase the reliability of operating machine. For machine-to-machine communication scenario, the SE-SC can efficiently support and benefit different machine communication mechanisms.

**Advantage 3: Flexible Semantic Resource Allocation**: The SE-SC can find the optimal number of sending semantic features and provide the variable-length semantic symbol transmission scheme for semantic resource allocation. For multiuser semantic communication scenario, the SE-SC is expected to flexibly allocate code-domain resources according to the number of required semantic symbols corresponding to different users.

## V. OUR PROPOSED SE-SC FOR SEMANTIC ANALYSIS TASK: A CASE

As a case, we investigate our proposed SE-SC to drive the downstream task. In this article, we consider the object detection task as a representative downstream task, but the proposed SE-SC is extensible for other downstream tasks such as classification, segmentation, and data reconstruction. In this study case, YOLOv5, which is widely used object detection model, is used as the downstream task model. In the SE-SC scheme, an adaptive rate transmission policy can be enabled by adjusting a non-trainable threshold value $\mu$ without training multiple models. Specifically, we will mask a subset of semantic features, where the weight of masked semantic feature is below the threshold value. Considering the provided data and the application, the weight vector highlights the importance of semantic feature in the downstream task model, thus the proposed rate transmission policy can dynamically select relatively important semantic features for sending to the receiver. Also, the number of transmitted semantic features can match the rate transmission rate by adjusting the threshold value. We compare our proposed SE-SC with the full semantic feature transmission and traditional codec scheme, which is based on the JPEG2000 image code combined with LDPC channel code. The slow Rayleigh fading channel is used to evaluate the robustness of the proposed SE-SC, along with other comparison schemes. We test our proposed SE-SC using the VOC 2012 dataset. The mean Average Precision (mAP), which is widely used in the object detection task, is adopted as the objective metric in this study case.

Figure 4 shows a visual example of our proposed SE-SC in object detection task, where we set Signal-to-Noise Ratio (SNR) as 15dB. It can be observed from Fig. 4 that the image patch in weight vector corresponding to the semantic feature with higher values represents more importance for the detection task. Based on the weight vector, our proposed SE-SC can further compress redundant semantic information while conveying the intended information. From Fig. 4, we can observe that our proposed SE-SC focuses on reconstructing the image region of detection objects while ignoring the real reconstruction of task-unrelated image region such as sky and street light. Thus, all cars can be properly detected according to the output image that preserves the intended semantic information.

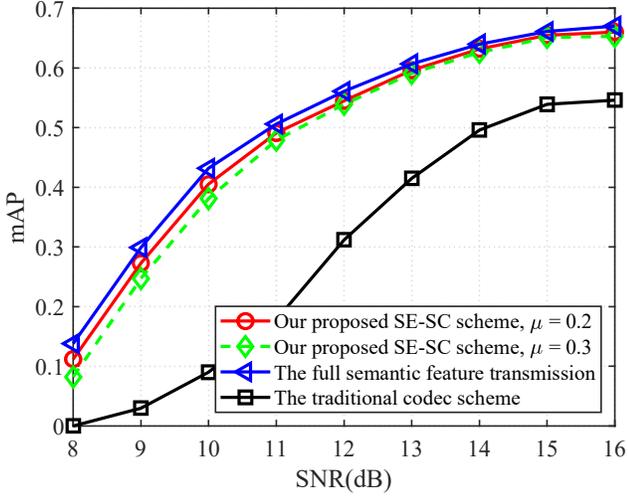

Fig. 5: mAP performance versus SNR, where the higher value means better performance.

Figure 5 shows mAP performance versus SNR at the slow Rayleigh fading channel. It can be observed from Fig. 5 that our proposed SE-SC scheme can achieve almost the same mAP as that of the full semantic feature transmission under the same SNR. It validates that the proposed SE-SC scheme can not only effectively offer an identical QoS for the downstream task but also save transmission symbols by compressing semantic feature. We can also observe from Fig. 5 that the proposed SE-SC scheme significantly outperforms the traditional codec scheme, especially in low SNR regions. Due to the sensitivity of traditional codec scheme to the bit error caused by channel noise, the traditional codec scheme will lose a lot of information carried by original data and thus its reconstructed data is difficult to be parsed to drive the downstream task. In high SNR regions, the traditional codec scheme aims to preserve data structure rather than semantic exchange, resulting in serious semantic distortion in critical data regions. It can be seen from Fig. 5 that as $\mu$ is adjusted from 0.2 to 0.3, the SE-SC scheme loses about 9.5% mAP in low SNR regions (SNR < 11dB) but saves about 23.7% transmission symbols. The mAP loss may be because some data only has few redundant semantic feature. In practice, $\mu$ can be initially set to be relatively low. If the number of transmitted semantic features fails to match the transmission rate, then we will gradually raise $\mu$ in order to reduce the number of transmitted semantic features until matching the transmission rate.

## VI. CHALLENGES FOR SCALABLE EXTRACTION BASED SEMANTIC COMMUNICATIONS

Despite the fact that the SE-SC shows many remarkable advantages, there still exist certain challenges associated with the implementation of SE-SC in the future 6G wireless networks.

In the following, we discuss several challenges and give some potential solutions.

**Challenge 1: The Exploitation of Lightweight SE-SC**: Considering the extremely massive connectivity for small devices in future Internet of Everything (IoE) networks, which is expected to use the SE-SC to improve network efficiency as well as support the intelligent connection. The SE-SC is with heavy neural network model to realize strong semantic parsing ability, thereby supporting partial semantic feature transmission. In practice, the SE-SC is heavy storage burden for most small devices, so it is highly demanded to exploit lightweight SE-SC. How to reduce the model complexity of the SE-SC while guaranteeing its capability for accurately recovering the data is a big challenge. The neural network pruning and quantization is an effective way to compress the size of the neural network model [12] and can be used for exploiting the lightweight SE-SC.

**Challenge 2: Combination of SE-SC and Channel Management**: Due to future 6G wireless network using advanced techniques such as Orbital Angular Momentum (OAM) and Reconfigurable Intelligent Surface (RIS), the channel modeling will become complex and thus it is very difficult to use the traditional statistics method. It is not only expected to use AI technology for semantic communication but also for channel management. The combination of SE-SC and channel management will give more freedom for semantic transmission. For example, SE-SC can use AI for channel modeling as well as channel estimation and then finds the optimal semantic feature compression ration to fit current wireless channel. For another example, AI-assisted RIS will enable the controlled radio environment in future networks [14]. With the programmable RIS, SE-SC can further improve the semantic compression limit by jointly optimizing the scalable semantic codec as well as the communication environment.

**Challenge 3: Tradeoff Between SE-SC Performance and Communication Overhead**: In fact, the SE-SC with superior performance relies on a shared basic knowledge with both transmitter and receiver as well as timely fine-tuning the SE-SC model. In practice, it is a extremely challenging to ensure that the basic knowledge of a large number of mobile users can be timely updated. Updating the basic knowledge of users over a wide area will result in significant communication overhead. Thus, it is crucially important to find optimal tradeoff between SE-SC performance and communication overhead for supporting the wide deployment of SE-SC. The mobile computing and edge service can be used to store the shared basic knowledge and fine-tune the SE-SC model, respectively. Then, the user can obtain the fine-tuned SE-SC model from the nearest edge server, which significantly reduces the communication overhead caused by sharing data over long distances.

**Challenge 4: Tradeoff Between Reconstruction Accuracy and Security**: Due to the fact that our proposed scalable semantic communication semantic can mask the private semantic features and the receiver relies on the corresponding basic knowledge to decode, it can also be regarded as a potential method for secure communications. However, using one shared feature representation for all private semantic fea-tures may result in severe reconstruction distortion, resulting in losing the private information at the legal receiver. Thus, we can train a codebook shared by the transmitter and the receiver to represent the semantic feature [15]. Then, the transmitter only sends the indices of semantic features in the codebook. However, we may need a large codebook to accurately represent all private semantic features.

## VII. Conclusion

In this article, we proposed a novel SE-SC model, which aims to send compressed semantic feature without losing the intended information. Then, we applied our proposed SE-SC framework to support the downstream tasks in 6G wireless networks. Simulation results have shown that the proposed SE-SC scheme can offer an identical QoS for the downstream task as the full semantic feature transmission and consumes much fewer transmission symbols. And, the SE-SC scheme significantly outperforms the traditional codec scheme. We also explored several challenges for further investigating the scalable extraction based semantic communications and demonstrated the considerable potential of developing 6G wireless networks.

## References

[1] ITU-R, "IMT for 2030 and beyond," in *Workshop*, 2023. [Online]. Available: https://www.itu.int/en/ITU-R/study-groups/rsg5/rwp5d/Pages/wsp-imt-vision-2030-and-beyond.aspx
[2] W. Weaver, "Recent contributions to the mathematical theory of communication," *The Mathematical Theory of Communication*, 1949.
[3] W. Yang, H. Du, Z. Q. Liew, W. Y. B. Lim, Z. Xiong, D. Niyato, X. Chi, X. Shen, and C. Miao, "Semantic communications for future internet: Fundamentals, applications, and challenges," *IEEE Communications Surveys & Tutorials*, vol. 25, no. 1, pp. 213–250, 2023.
[4] Z. Weng and Z. Qin, "Semantic communication systems for speech transmission," *IEEE Journal on Selected Areas in Communications*, vol. 39, no. 8, pp. 2434–2444, Aug. 2021.
[5] H. Xie, Z. Qin, G. Y. Li, and B. H. Juang, "Deep learning enabled semantic communication systems," *IEEE Transactions on Signal Processing*, vol. 69, pp. 2663–2675, 2021.
[6] E. Bourtsoulatze, D. Burth Kurka, and D. Gündüz, "Deep joint source-channel coding for wireless image transmission," *IEEE Transactions on Cognitive Communications and Networking*, vol. 5, no. 3, pp. 567–579, 2019.
[7] K. He, X. Zhang, S. Ren, and J. Sun, "Deep residual learning for image recognition," in *2016 IEEE Conference on Computer Vision and Pattern Recognition (CVPR)*, 2016, pp. 770–778.
[8] M. Li, W. Zuo, S. Gu, D. Zhao, and D. Zhang, "Learning convolutional networks for content-weighted image compression," in *2018 IEEE/CVF Conference on Computer Vision and Pattern Recognition*, 2018, pp. 3214–3223.
[9] J. Dai, S. Wang, K. Tan, Z. Si, X. Qin, K. Niu, and P. Zhang, "Nonlinear transform source-channel coding for semantic communications," *IEEE Journal on Selected Areas in Communications*, vol. 40, no. 8, pp. 2300–2316, 2022.
[10] K. He, X. Chen, S. Xie, Y. Li, P. Dollr, and R. Girshick, "Masked autoencoders are scalable vision learners," in *2022 IEEE/CVF Conference on Computer Vision and Pattern Recognition (CVPR)*, 2022, pp. 15 979–15 988.
[11] D. Huang, F. Gao, X. Tao, Q. Du, and J. Lu, "Toward semantic communications: Deep learning-based image semantic coding," *IEEE Journal on Selected Areas in Communications*, vol. 41, no. 1, pp. 55–71, 2023.
[12] H. Xie and Z. Qin, "A lite distributed semantic communication system for internet of things," *IEEE Journal on Selected Areas in Communications*, vol. 39, no. 1, pp. 142–153, Jan. 2021.
[13] Z. Chen and T. He, "Learning based facial image compression with semantic fidelity metric," *Neurocomputing*, vol. 338, pp. 16–25, 2019.
[14] H. Gacanin and M. Di Renzo, "Wireless 2.0: Toward an intelligent radio environment empowered by reconfigurable meta-surfaces and artificial intelligence," *IEEE Vehicular Technology Magazine*, vol. 15, no. 4, pp. 74–82, 2020.
[15] Q. Hu, G. Zhang, Z. Qin, Y. Cai, G. Yu, and G. Y. Li, "Robust semantic communications with masked vq-vae enabled codebook," *IEEE Transactions on Wireless Communications*, pp. 1–1, 2023.

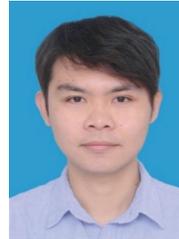

**Yuzhou Fu** received the M.S. degree in school of computer, electronics and information from Guangxi University, Guangxi, China, in 2019. He is currently pursuing a Ph.D. degree in telecommunication engineering at Xidian University. His research interests include semantic communication and maritime wireless communication network optimization.

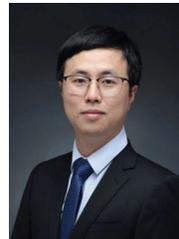

**Wenchi Cheng** (M'14-SM'18) received his B.S. and Ph.D. degrees in telecommunication engineering from Xidian University in 2008 and 2013, respectively, where he is a full professor. He was a visiting scholar with the Department of Electrical and Computer Engineering, Texas A&M University, College Station, from 2010 to 2011. His current research interests include B5G/6G wireless networks, emergency wireless communications, and OAM based wireless communications. He has published more than 100 international journal and conference papers in the IEEE JSAC, IEEE magazines, and IEEE transactions, and at conferences including IEEE INFOCOM, GLOBECOM, ICC, and more. He has served or is serving as an Associate Editor for the IEEE Systems Journal, IEEE Communications Letters, and IEEE Wireless Communications Letters, as the Wireless Communications Symposium Co-Chair for IEEE ICC 2022 and IEEE GLOBECOM 2020, the Publicity Chair for IEEE ICC 2019, and the Next Generation Networks Symposium Chair for IEEE ICCC 2019.

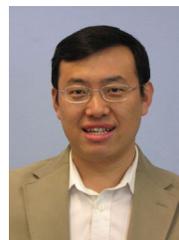

**Wei Zhang** (S'01-M'06-SM'11-F'15) is a Professor at the School of Electrical Engineering and Telecommunications, the University of New South Wales, Sydney, Australia. His current research interests include 6G communications and networks. He is Vice President of IEEE Communications Society. He also serves as an Area Editor of the IEEE Transactions on Wireless Communications and the Editor-in-Chief of Journal of Communications and Information Networks.

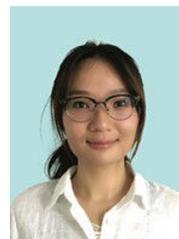

**Jingqing Wang** received the B.S. degree from Northwestern Polytechnical University, Xi'an, China, in Electronics and Information Engineering and the Ph.D. degree in Electrical and Computer Engineering from Texas A&M University, College Station, TX, USA. She is currently an Assistant Professor at Xidian University. She won the Best Paper Award from the IEEE GLOBECOM in 2020 and 2014, respectively, the Hagler Institute for Advanced Study Heep Graduate Fellowship Award from Texas A&M University in 2018, and Dr. R.K. Pandey and Christa U. Pandey'84 Fellowship, Texas A&M University, USA, 2020-2021. Her research interests focus on B5G/6G mobile wireless network technologies, statistical QoS provisioning, 6G mURLLC, information-theoretic analyses of FBC, and emerging machine learning techniques over 5G and beyond mobile wireless networks.